\documentclass[12pt]{iopart}
\pdfoutput=1
\usepackage{graphicx}
\usepackage{fancyhdr}
\usepackage{amssymb,amsfonts,color}

\definecolor{red}{rgb}{1,0,0}

\begin{document}

\title[]{Spin networks on adiabatic quantum computer}

\author{Jakub Mielczarek}

\address{Institute of Physics, Jagiellonian University, 
{\L}ojasiewicza 11, 30-348 Krak\'{o}w, Poland}

\begin{abstract}
The article is addressing a possibility of implementation of spin network states 
on adiabatic quantum computer. The discussion is focused on application of 
currently available technologies and analyzes a concrete example of D-Wave 
machine. A class of simple spin network states which can be implemented on the 
Chimera graph architecture of the D-Wave quantum processor is introduced. 
However, extension beyond the currently available quantum processor 
topologies is required to simulate more sophisticated spin network 
states, which may inspire development of new generations of adiabatic 
quantum computers. A possibility of simulating Loop Quantum Gravity
is discussed and a method of solving a graph non-changing scalar 
(Hamiltonian) constraint with the use of adiabatic quantum computations 
is proposed. 
\end{abstract}

\maketitle

\section{Introduction}

One can distinguish two main modes of operation of quantum computers. 
The first is quantum data processing associated mostly with implementation 
on quantum algorithms with the use of quantum gates or various quantum 
machine learning protocols \cite{Biamonte}. The second concerns simulations 
of quantum systems. 

Simulating quantum system with the use quantum computers is fundamentally 
different from what simulations performed at classical computers are. While 
classical simulations rely on either discretization of a given physical system 
or an adequate algebraic analysis the simulations performed on quantum 
computers allow to \emph{imitate} a given quantum systems. This kind of 
\emph{exact simulation} of a quantum system has been a subject of discussion 
in a seminal R.~Feynman article \cite{Feynman:1981tf}.  

In order to understand better what we mean by exact simulations let us consider 
the case of Planck scale physics. Here, the relevant degrees of freedom are 
defined at length scales of the order of the Planck length $l_{\rm Pl} \sim 10^{-35}$m.
Despite significant advances made in both theoretical understanding and 
experimental techniques, the Planck scale physics remains empirically 
inaccessible for the moment.

\begin{figure}[ht!]
\centering
\includegraphics[width=10cm,angle=0]{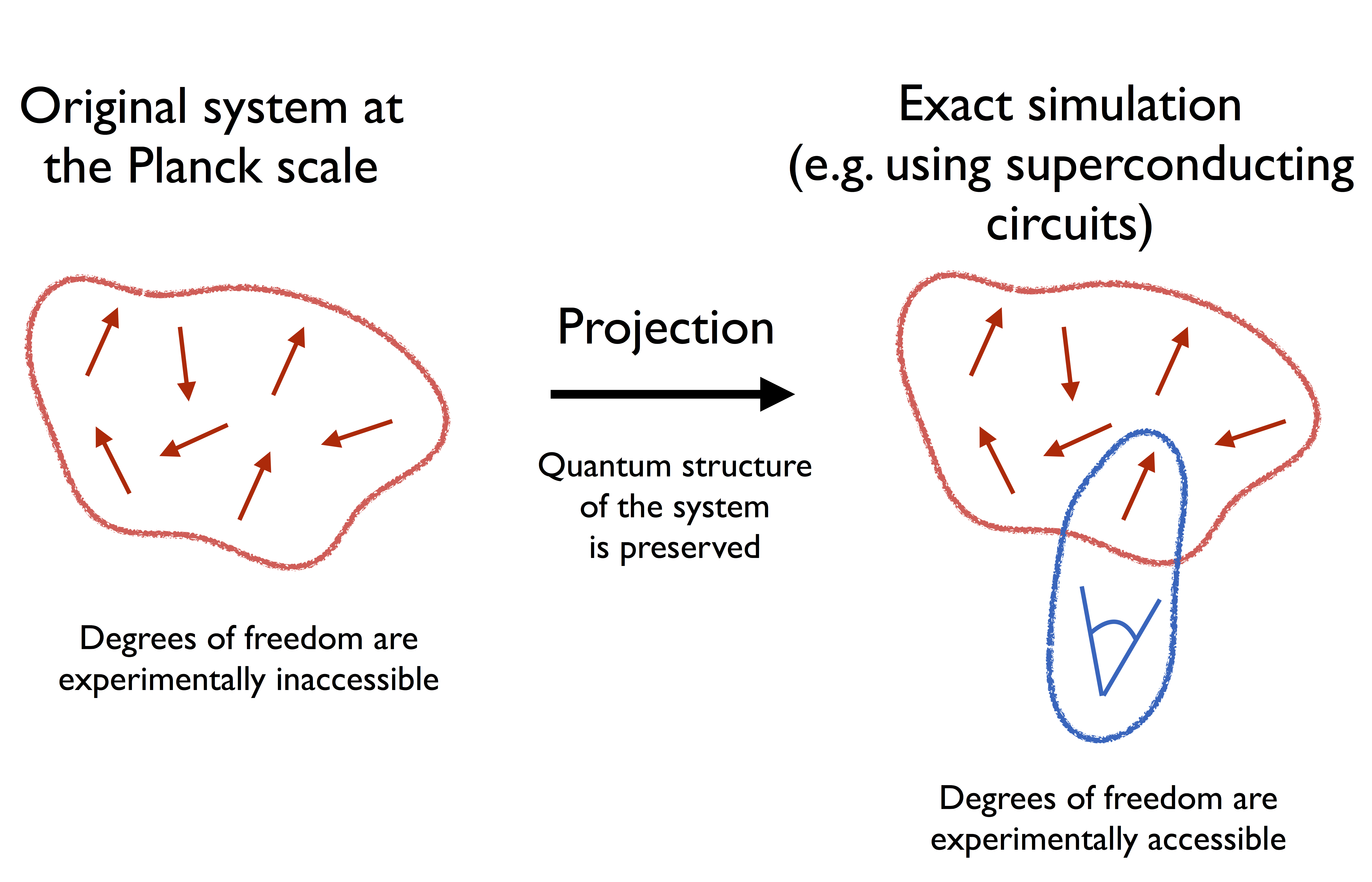}
\caption{Pictorial presentation of the relation between the original 
quantum system defined at the Planck scales and its exact simulation. 
The exact simulation is performed with the use of projection of the 
original quantum system onto the architecture of a quantum computer. 
In contrast to the Planck scale system, measurements of the quantum 
degrees of freedom can be performed at the level of the quantum simulation.} 
\label{Projection}
\end{figure}

On the other hand, concrete examples of theories describing elementary 
quantum gravitational degrees of freedom exist. One such approach is 
Loop Quantum Gravity (LQG) \cite{Ashtekar:2004eh}. In LQG, background 
independent degrees of freedom can be defined and some predictions 
can be made \cite{Barrau:2013ula}. Even if the degrees of freedom under 
consideration are experimentally not accessible one can think about their 
\emph{projection} onto another physical realization which will imitate its 
quantum behavior (see Fig. \ref{Projection}). Assuming that quantum mechanics 
is valid at the Planck scale, from the perspective of quantum theory such 
systems can be considered as equivalent. The only difference are appropriately 
rescalled couplings adjusted to the physical nature of the simulator (built e.g. 
with the use of superconducting qubits). Such projection of one quantum system 
into its equivalent imitation allows to perform what we previously called exact 
simulations. As we already mentioned, the quantum simulations are very 
different from what we usually consider as physical simulations, where for 
instance a given differential equation is discretized and then implemented 
on a computer with the use of appropriate algorithms. In contrast, in the case 
of exact simulations one actually does experiments on a quantum system 
which is defined as being equivalent (from the view point of quantum mechanics) 
to a part (or a whole) of the original quantum system. 

The aim of the is article is to investigate a possibility of performing exact 
quantum simulations of the spin networks states which are used to construct 
Hilbert space of LQG approach to quantum gravity. We are interested in 
application of existing technology, therefore our focus is on the only  
commercially available quantum computer at the moment, namely the D-Wave 
machine which implements the so-called quantum annealing algorithm. 
 
Furthermore, we are considering conceptual issues related with the possibility 
of simulating Loop Quantum Gravity based on the considered fixed graph 
case. Specifically, implementation of the scalar constraint is analyzed and 
a toy model of such a procedure is presented. We finish with an outlook of the next
steps to be done in the research direction initiated in this article.  

Worth mentioning here is that the idea of employing the spin networks states 
in quantum computations already appeared in the literature (see Refs. \cite{MR,Jordan,Kauffman}). 
However, the potential of application of spin networks for the purpose of universal 
quantum data processing was considered only. Up to the best of our knowledge the 
issue of relating spin networks with adiabatic quantum computations was 
not considered before. Furthermore, while this article was in the final stage of 
preparation a study in which a LQG spin network is implemented on a molecular 
quantum simulator appeared. In the article a simulation of quantum fluctuations 
of a 5-node spin network in the kinematical regime was performed \cite{Li:2017gvt}. 
Here, we will consider the same type of spin network in Sec. \ref{ImpLQG} in the 
context of solving a prototype scalar constraint with the use of adiabatic quantum 
computations. 

\section{Adiabatic Quantum Computing}

The last years have brought a significant progress in the development of 
quantum computing technologies \cite{Campbell}. First quantum computers 
have been commercialized and made available in a cloud or as an independent 
hardware units. In both cases the currently most advanced commercial 
technologies were possible to achieve thanks to the development of 
superconducting quantum circuits \cite{You}. In particular, the IBM Q universal quantum 
computer built with the use of 5 and 20 superconducting qubits has been 
developed. However, from the point of view of exact simulations discussed 
in the Introduction, another type of quantum computer seems to be more 
suitable to use - namely the \emph{adiabatic quantum computer} \cite{Kadowaki}. 

The adiabatic quantum computers, in contrast to the universal ones, are 
designed to solve a specific problem of finding minimum of a Hamiltonian 
$H_{I}$ of a coupled system of qubits (spins). In the process of finding the 
minimum of $H_{I}$ one employs a time dependent Hamiltonian in the form 
\begin{equation}
H(\lambda) = (1-\lambda)H_{B}+\lambda H_{I},
\end{equation}
where $H_{B}$ is so-called base Hamiltonian which is characterized by a 
simple and easy to prepare ground state. In practice, the base Hamiltonian is
often equal to $H_{B} = \sum_{i} \sigma^{x}_i$, such that the ground state 
corresponds to the alignment of spins in the $x$ direction. Then, the value of 
$\lambda$ is changed adiabatically from $\lambda=0$ to $\lambda = 1$, such 
that while the system is initially in non-degenerate ground state it will remain 
in a ground state. Therefore, if the process is done correctly, the system ends 
up in the minimum of the Hamiltonian $H_{I}$. The process of transition from 
$\lambda=0$ to $\lambda = 1$ involves quantum tunneling and is called 
\emph{quantum annealing}. The characteristic time scales which preserve 
adiabaticity are dependent on what kind of $H_{I}$ Hamiltonian is 
considered. The issue is closely related with the efficiency of quantum annealing
based algorithms with respect to the classical ones (see Ref. \cite{Biamonte} for 
discussion of this subject). 
  
In practical implementations the most considered form of $H_{I}$ corresponds 
to the Ising problem:
\begin{equation}
H_{I} = \sum_{<i,j>} b_{ij} \sigma^{z}_i\sigma^{z}_i + \sum_i h_i \sigma^{z}_i, \label{HI}
\end{equation}
where $b_{ij}$ are coupling between spins and $h_{i}$ quantifies interactions
of spins with external magnetic field.  The summation $<i,j>$  is defined such that 
it does not repeat over pairs. The values of couplings define the problem to be 
solved while readout of $z$ components of the spins in the final state provides an 
outcome of the quantum computation (simulation). 

From the mathematical viewpoint, the class of problems which can be solved 
in that way is the so-called Quadratic Unconstrained Binary Optimization (QUBO) 
which typically is of the NP hard type.  This is because, when we look a the problem 
from the classical perspective by measuring the orientations of spins along the 
$z$-axis, the two values of $\sigma^{z}_i \rightarrow s_i\in \{ -1, 1 \} $ are allowed. 
In consequence, for the system of $N$ classical spins, there are $2^{N}$ configurations
to be explored. Therefore, in general, finding a ground state requires exponential 
growth of time with the number of spins ($N$).   

The physical implementation of the QUBO problem with the use of quantum 
annealing procedure is provided by the D-Wave machine. In this realization, 
the spins (qubits) are created with the use of superconducting circuits in the 
form of CC JJ RF-SQUIDs \cite{Harris} built with the use of Josephson junctions composed 
of Niobium in superconducting state. The qubit base states are defined employing  
two different orientations of quantum of magnetic flux across the superconducting 
circuit. Interactions between the qubits are introduced by SQUID (superconducting 
quantum interference device) based circuits called \emph{couplers}, which 
introduce the $b_{ij} \sigma^{z}_i\sigma^{z}_i$ factors in the Hamiltonian (\ref{HI}). 
Furthermore, with the use of external magnetic fluxes the values of parameters $h_i$ can be 
controlled. However, not all values of $b_{ij}$ and $h_i$ are allowed but only some 
fractional values from the range $[-1,1]$. The readout of the final 
quantum states of qubits is performed with the use of sensitive magnetometers 
built with the use of SQUIDs.  

\begin{figure}[ht!]
\centering
\begin{tabular}{ccc}
a) \includegraphics[width=5cm,angle=0]{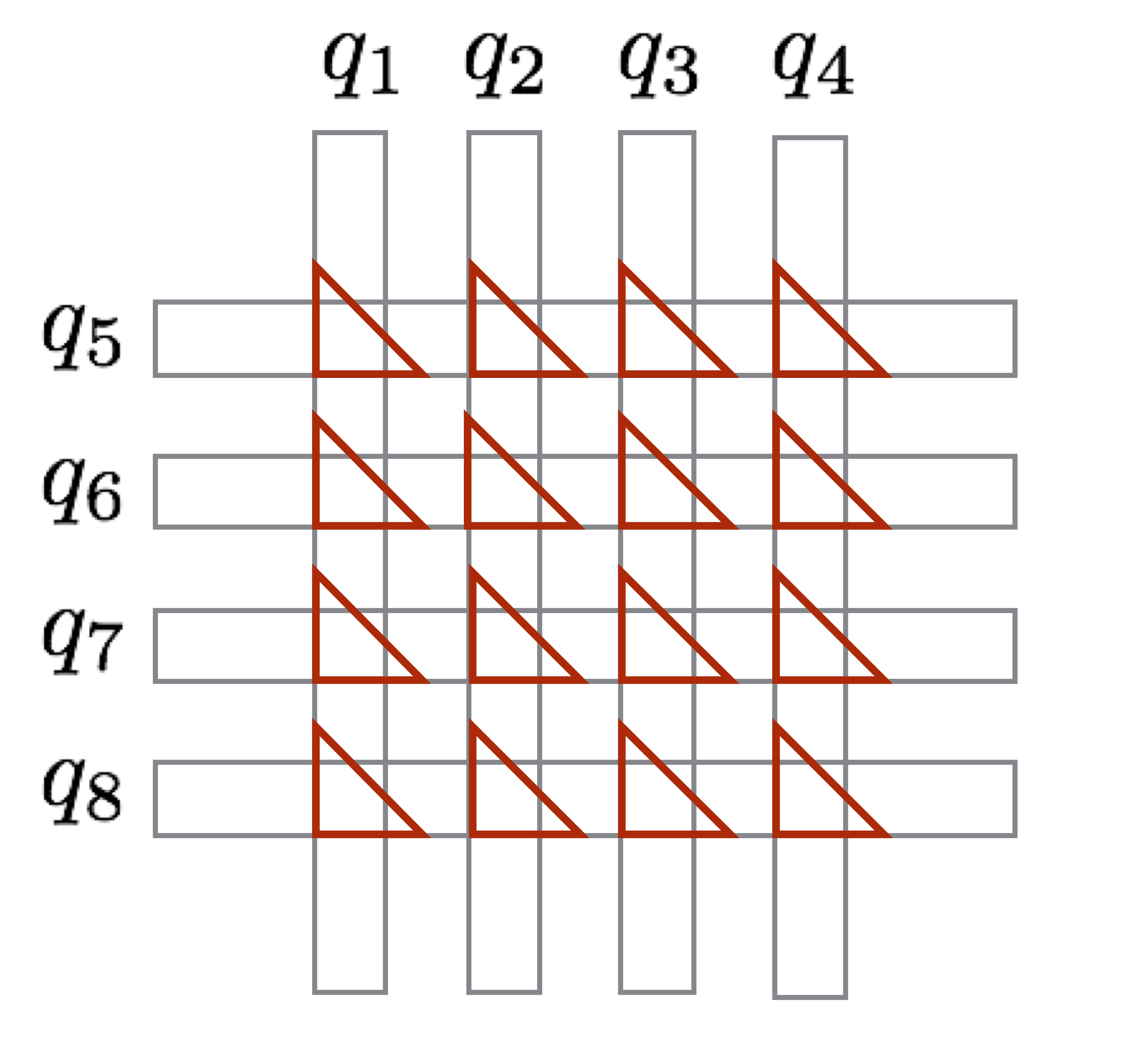} &
b) \includegraphics[width=4cm,angle=0]{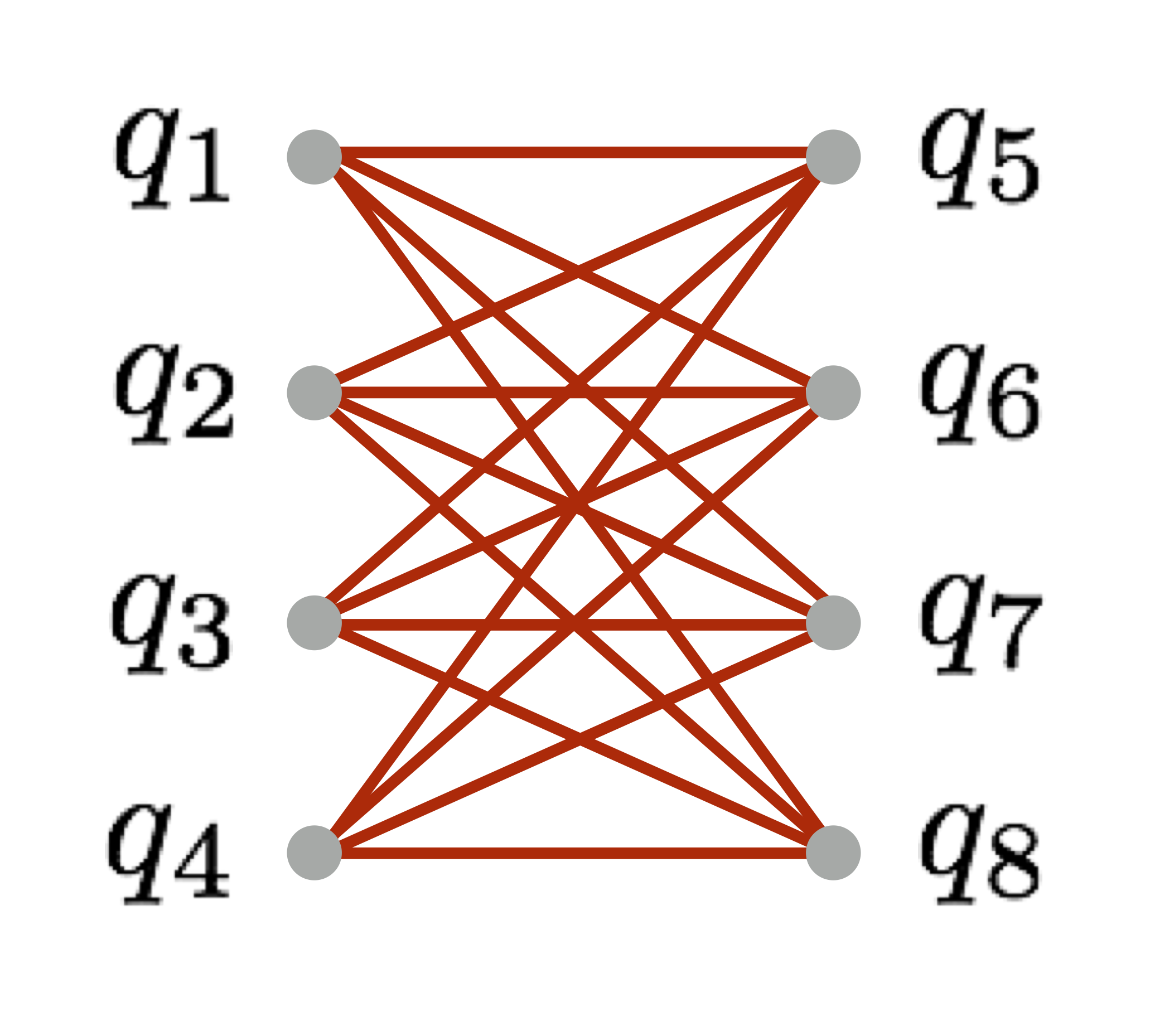} &
c) \includegraphics[width=4cm,angle=0]{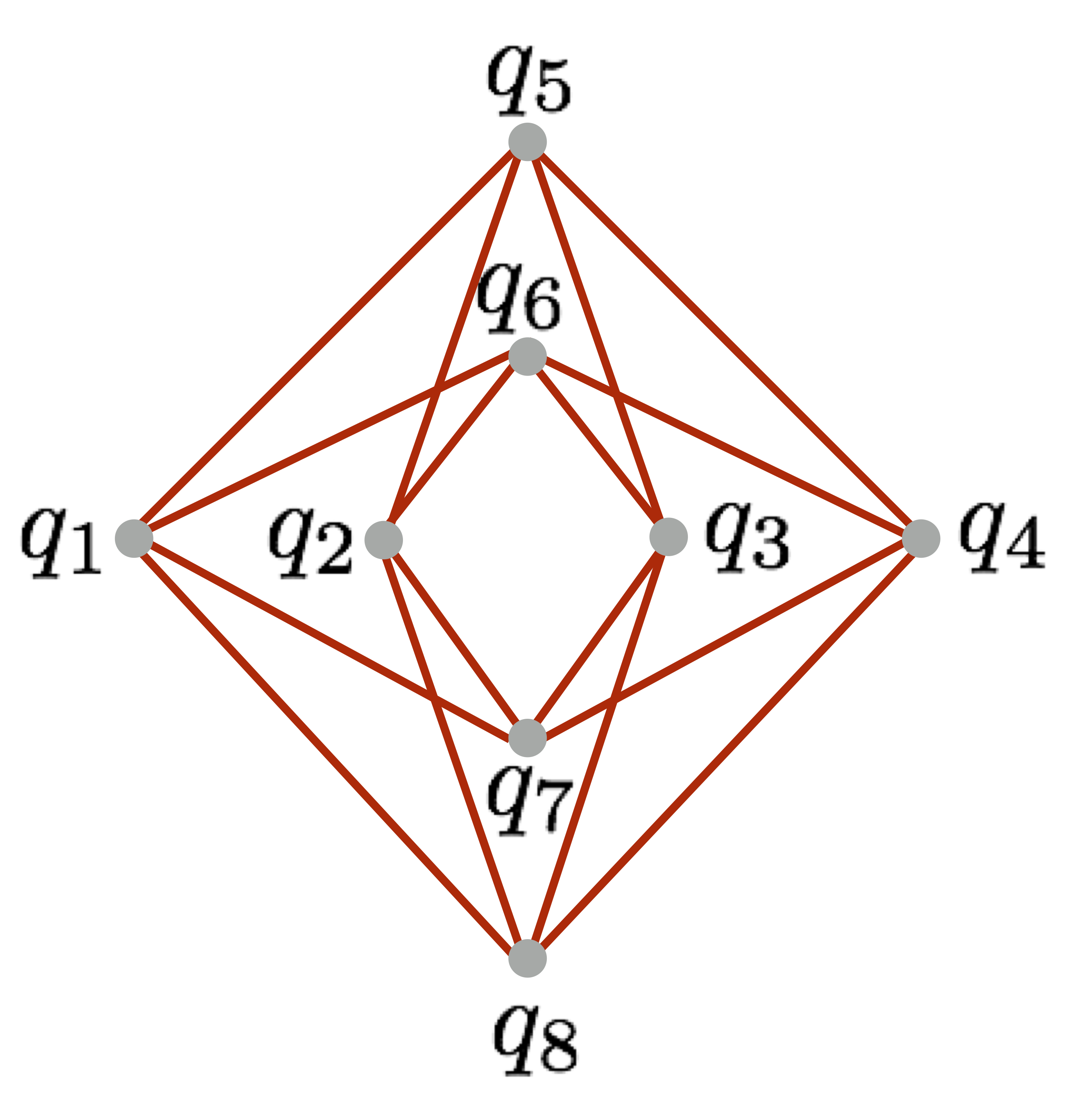}
\end{tabular}
\caption{Three different representations of the structure of couplings
between eight $q_i$ qubits forming an elementary block of the D-Wave 
processor: a) The physical representation of quibits as closed superconducting 
loops. The couplers between the qubits are represented by triangles. b) The Chimera 
graph structure of couplings between the eight qubits. c) Representation of a single 
block which is useful when interconnections in the array of blocks are considered.} 
\label{Chimera}
\end{figure}

In the D-Wave quantum annealer, the superconducting qubits $q_i$ are arranged 
into 8-qubit blocks forming the so-called \emph{Chimera} architecture. Each block 
has 16 couplings between 8 spins. Therefore, not all qubits are coupled. The topology of 
couplings between qubits in a single block is presented in Fig. \ref{Chimera}.
In the so far most advanced version of the D-Wave machine (the D-Wave 2000) 
the 8-qubit blocks form a 16x16 matrix (256 blocks in total) leading to 2048 qubits.

\section{Spin Networks}

Let us now move to the subject of spin networks. We will start with a brief 
overview of how spin networks appear in the Loop Quantum Gravity approach 
to quantum gravity. Then we will introduce a class of spin networks which 
is possible to implement on Chimera architecture of a quantum processor. 

The fundamental elements of Loop Quantum Gravity approach to quantum gravity are 
holonomies of Ashtekar connection $A \in su(2)$ along some curve $e(\lambda)$ 
with $\lambda \in [0,1]$:
\begin{equation}
h[A,e] = \mathcal{P} e^{\int_e A}. 
\end{equation}
Performing gauge transformations, generated by the so-called Gauss constraint, 
the Ashtekar connection transforms as $A\rightarrow A_g = g^{-1}dg+g^{-1}Ag$.
The corresponding transformation of holonomy is $h[A,e] \rightarrow h[A_g,e] =g(e(0))h[A,e]g(e(1))^{-1}$. 
The fact that the transformations of holonomies contribute only at the boundaries of $e$ implies that 
gauge invariant objects are provided by the Wilson loops $W[A,e] :={\rm tr}\left(h[A,e] \right)$.
 
The key idea behind LQG is to built a Hilbert space of the theory out of the 
Wilson loops. However, such basis is in general over-complete.  A  
solution to the problem comes from construction of \emph{spin-networks} which 
are certain linear combination of products of the Wilson loops \cite{Rovelli:1989za}. 
Such approach guarantees that both the Gauss constraint (ensuring local gauge invariance) 
is satisfied by the base states and the Hilbert space is complete. Furthermore, 
by introducing equivalence relation between topologically equivalent spin networks, 
the so-called diffeomorphism constraint can be satisfied.  There is finally a 
scalar constraint which has to be satisfied by physical states. In this section 
we will focus on the spin networks states satisfying both the Gauss and diffeomorphism 
constraint. We will came back to the issue of satisfying the scalar constraint
(with the use of adiabatic quantum computing) in the next section. 

The spin network is formally a graph composed of edges $E$ and nodes $N$ with spin labels 
at the edges and so-called \emph{intertwiners} at the nodes. The spin labels 
correspond to irreducible representations of the $SU(2)$ group such that triangle 
inequalities (reflecting the Gauss constraint) are satisfied at the nodes. The intertwiners 
correspond to invariant subspaces at the nodes, which we will discuss in more details below. 

\begin{figure}[ht!]
\centering
\begin{tabular}{cc}
a)\includegraphics[width=6cm,angle=0]{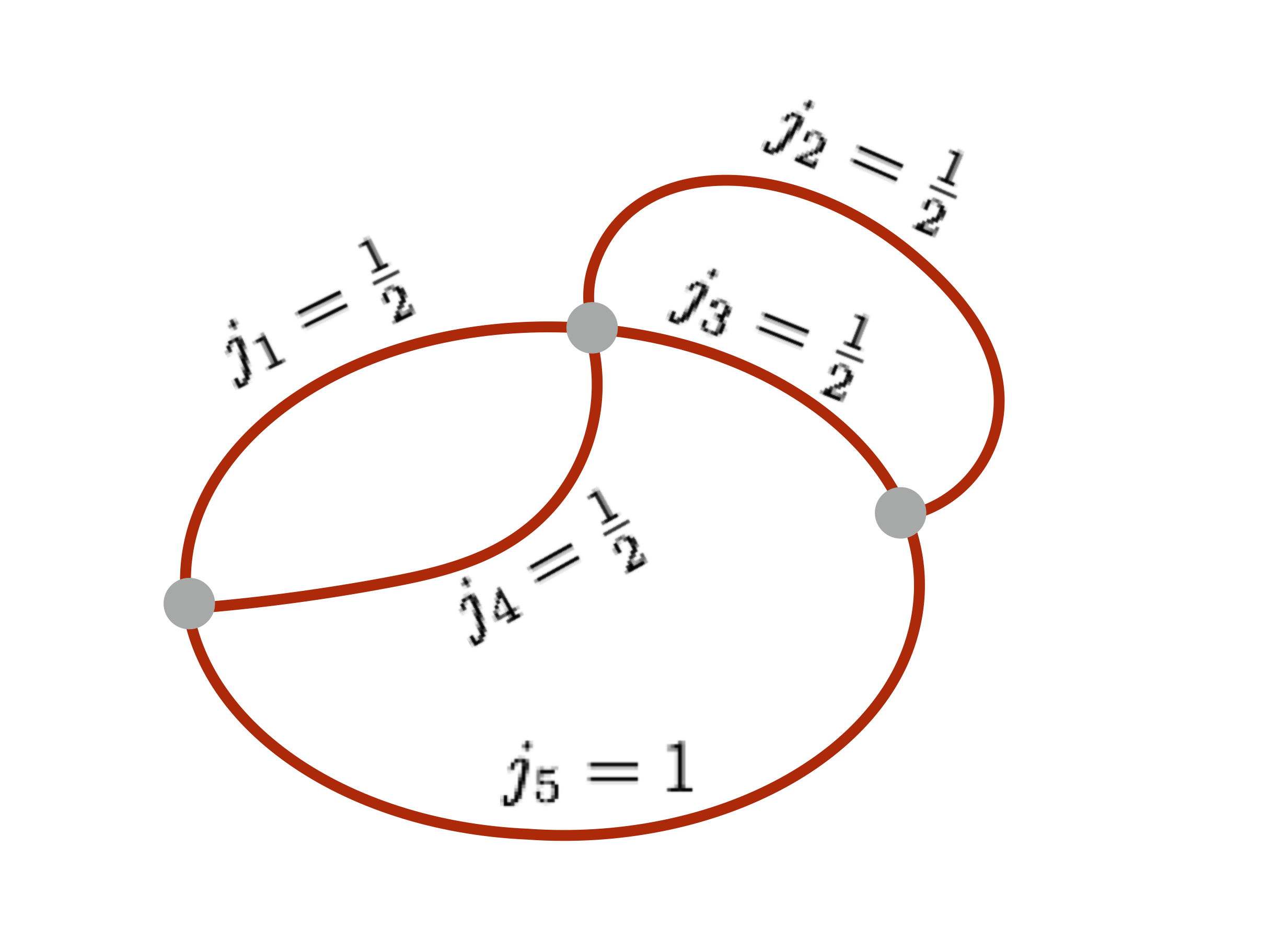} &
b)\includegraphics[width=8cm,angle=0]{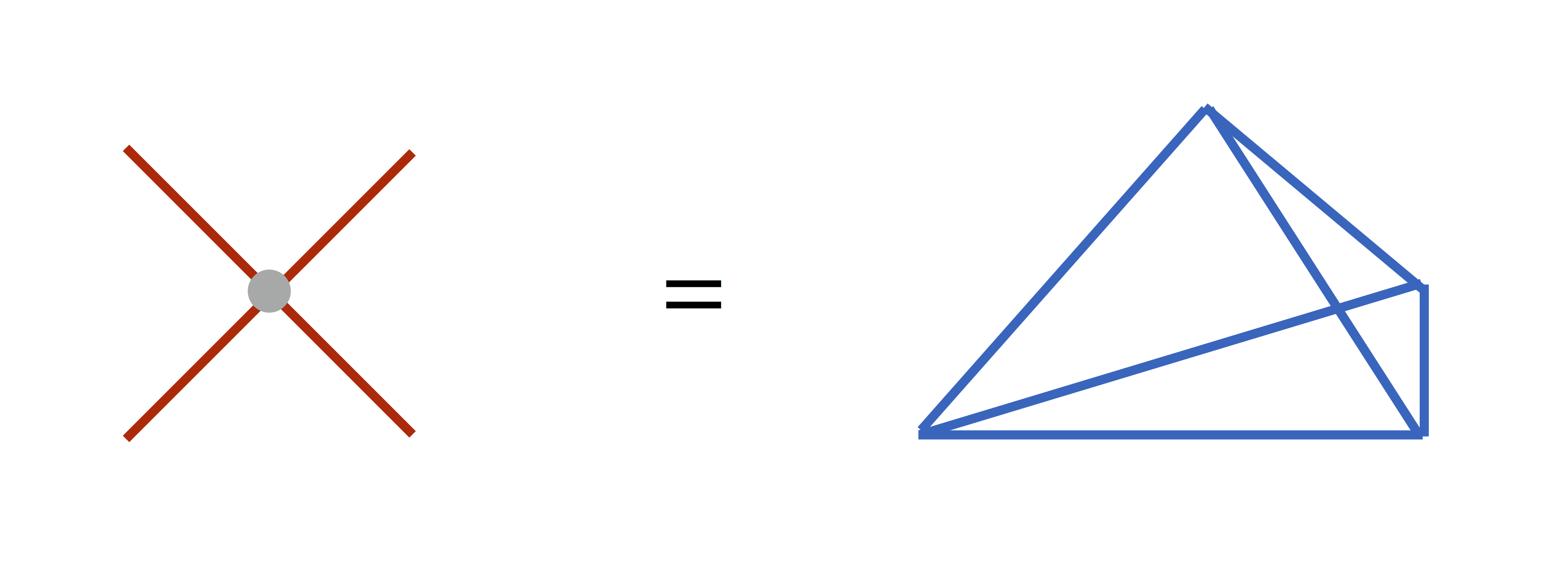}
\end{tabular}
\caption{a) An exemplary spin network. b) In the geometric picture the 4-valent node is dual 
to a tetrahedron (3-simplex).} 
\label{SpinNet}
\end{figure}

In Fig. \ref{SpinNet} we present an exemplary spin network composed of 
3-valent and 4-valent nodes. An important feature of the nodes is that 
3-valent  nodes do not carry a volume element while 4-valent nodes 
and higher valent nodes are associated with 3-volume (in the sense that 
eigenvalues of the volume operator in such a state are non-vanishing). 

The special case of the nodes are  4-valent nodes which in the geometric picture are 
dual to a tetrahedra (3-simplexes). One can imagine that the vertex is located in the center
of the tetrahedra  while each of the associated links intersects with one of the surfaces 
(see block b) in Fig. \ref{SpinNet}). 

\begin{figure}[ht!]
\centering
\includegraphics[width=10cm,angle=0]{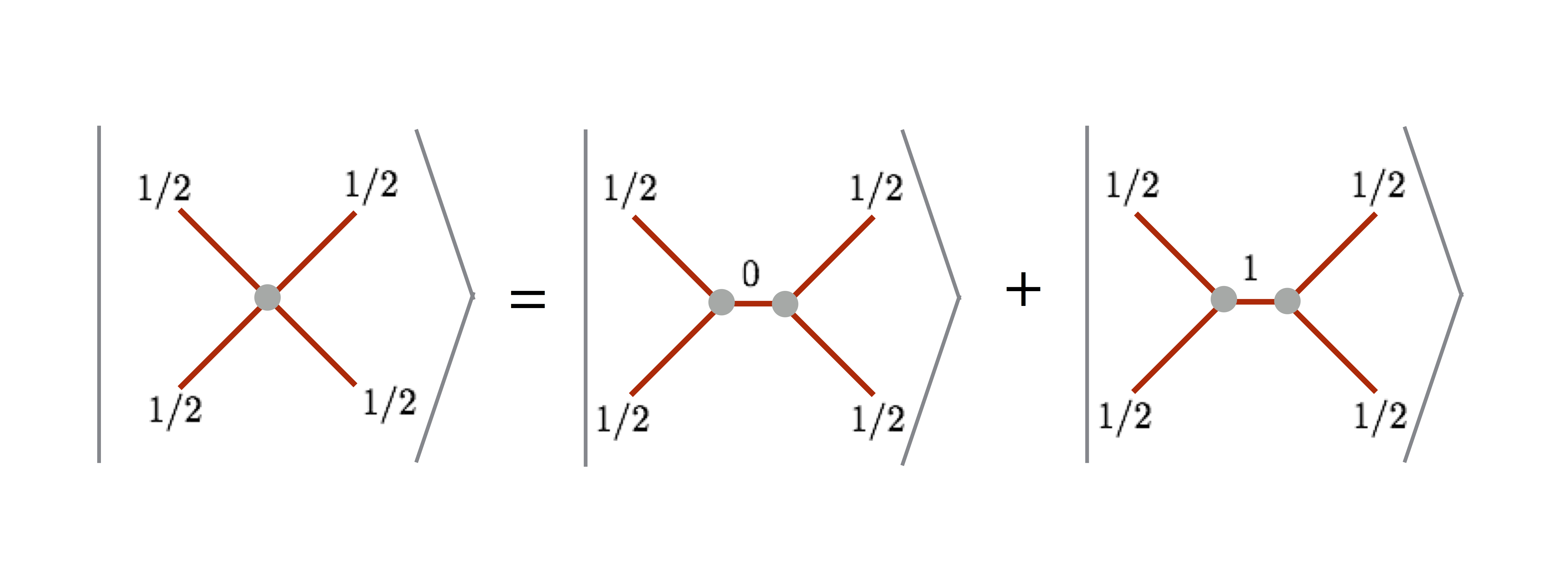}
\caption{The 4-valent node with spin labels equal to $j=1/2$ corresponds to superposition 
of two graphs each with the different value of intertwiner. In this article, the two dimensional 
intertwiner space is associated with the qubit Hilbert space.} 
\label{Qubit4V}
\end{figure}

In this article we are considering the case of spin networks composed only out 
of 4-valent vertices and spin labels corresponding to fundamental representations of 
the $SU(2)$ group i.e. $j=1/2$. The reason for that is that in such a case the Hilbert 
space at each vertex is a tensor product of four 1/2 spins which can be decomposed 
into irreducible representations in the following way:
\begin{eqnarray}
H_{1/2}\otimes H_{1/2}\otimes H_{1/2}\otimes H_{1/2} 
= H_{0} \oplus H_{0}  \oplus 3 H_{1}  \oplus H_2. 
\end{eqnarray}
There are, therefore, two possibilities in which the spins can add up 
to zero. In consequence, the invariant subspace for such a vertex is two 
dimensional: 
\begin{equation}
{\rm dim\ Inv} (H_{1/2}\otimes H_{1/2}\otimes H_{1/2}\otimes H_{1/2})=2.
\end{equation}
We associate the two dimensional invariant space with the qubit space. 
The nature of the qubit associated with the 4-vertex under consideration is 
graphically presented in Fig. \ref{Qubit4V}. 
Worth mentioning here is that there is a freedom of choice of channel in which 
recoupling of spin labels at the vertex is performed. In particular, in the $s$ 
channel the base states can be expressed in terms of the four $1/2$ spin 
base states ($|\uparrow \rangle$ and $|\downarrow \rangle$ ) as follows \cite{Feller:2015yta}:
\begin{eqnarray}
| 0_s \rangle &=& \frac{1}{2}\left(|\uparrow\downarrow\uparrow\downarrow\rangle
+|\downarrow\uparrow\downarrow\uparrow\rangle 
-|\uparrow\downarrow\downarrow\uparrow\rangle 
-|\downarrow\uparrow\uparrow\downarrow\rangle   \right) \nonumber \\
| 1_s \rangle &=& \frac{1}{\sqrt{3}} \left( |\uparrow\uparrow\downarrow\downarrow\rangle
+|\downarrow\downarrow\uparrow\uparrow\rangle
-\frac{|\uparrow\downarrow\uparrow\downarrow\rangle
+|\downarrow\uparrow\downarrow\uparrow\rangle 
+|\uparrow\downarrow\downarrow\uparrow\rangle 
+|\downarrow\uparrow\uparrow\downarrow\rangle}{2} \right)\nonumber. \\
\label{schannel}
\end{eqnarray}
Then, it is convenient to construct our qubit states such that they are 
eigenvectors of the volume operator $\hat{V}$ (see e.g. Ref. \cite{Feller:2015yta} for 
details). This can be achieved  by considering the following superpositions 
of the states (\ref{schannel}): 
\begin{eqnarray}
|1\rangle &=& \frac{1}{\sqrt{2}}\left(| 0_s \rangle - i | 1_s \rangle  \right), \\ 
|0\rangle &=& \frac{1}{\sqrt{2}}\left(| 0_s \rangle + i | 1_s \rangle  \right),
\end{eqnarray} such that 
$\hat{V}|1\rangle= +V_0|1\rangle$ and $\hat{V}|0\rangle= -V_0|0\rangle$, where
$V_0 = \frac{\sqrt{3}}{4} l^3_{Pl}$ is the absolute value of the quantum of volume. 
The $|1\rangle$ and $|0\rangle$ are the qubit states that we refer to in the rest 
of this article. From the geometrical point of view, the two base states are associated 
with two orientations of the elementary 3D symplex of space. Worth stressing is 
that the volume can be either positive (for $|1\rangle$) or negative  (for $|0\rangle$). 
Therefore, in LQG the elementary volume can contribute with both signs. However, 
it is expected that in the semiclassical limit one of the contribution will dominate 
such that the net configuration will be characterized by non-vanishing volume.     

\begin{figure}[ht!]
\centering
\begin{tabular}{cc}
a) \includegraphics[width=7cm,angle=0]{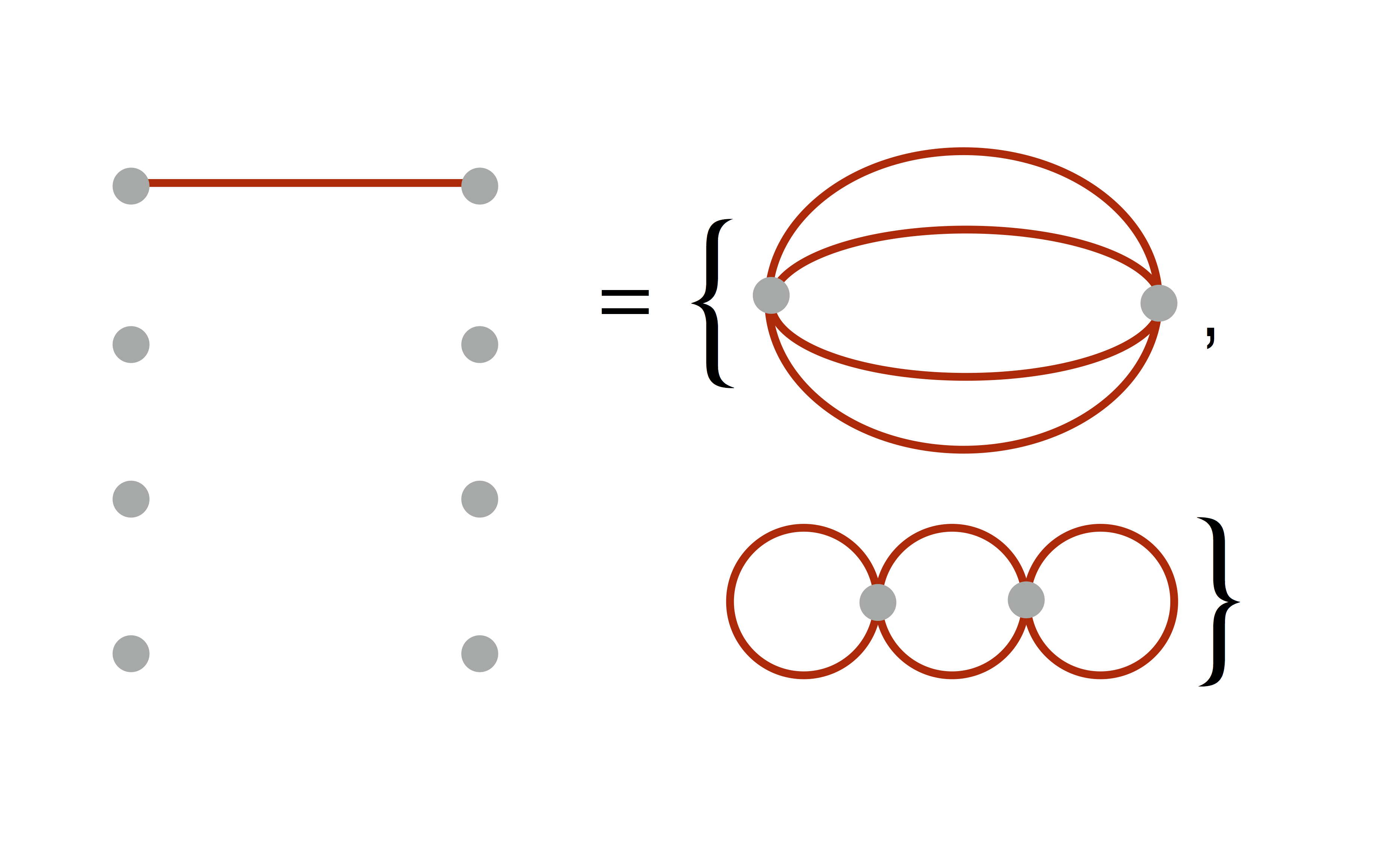} & 
b) \includegraphics[width=7cm,angle=0]{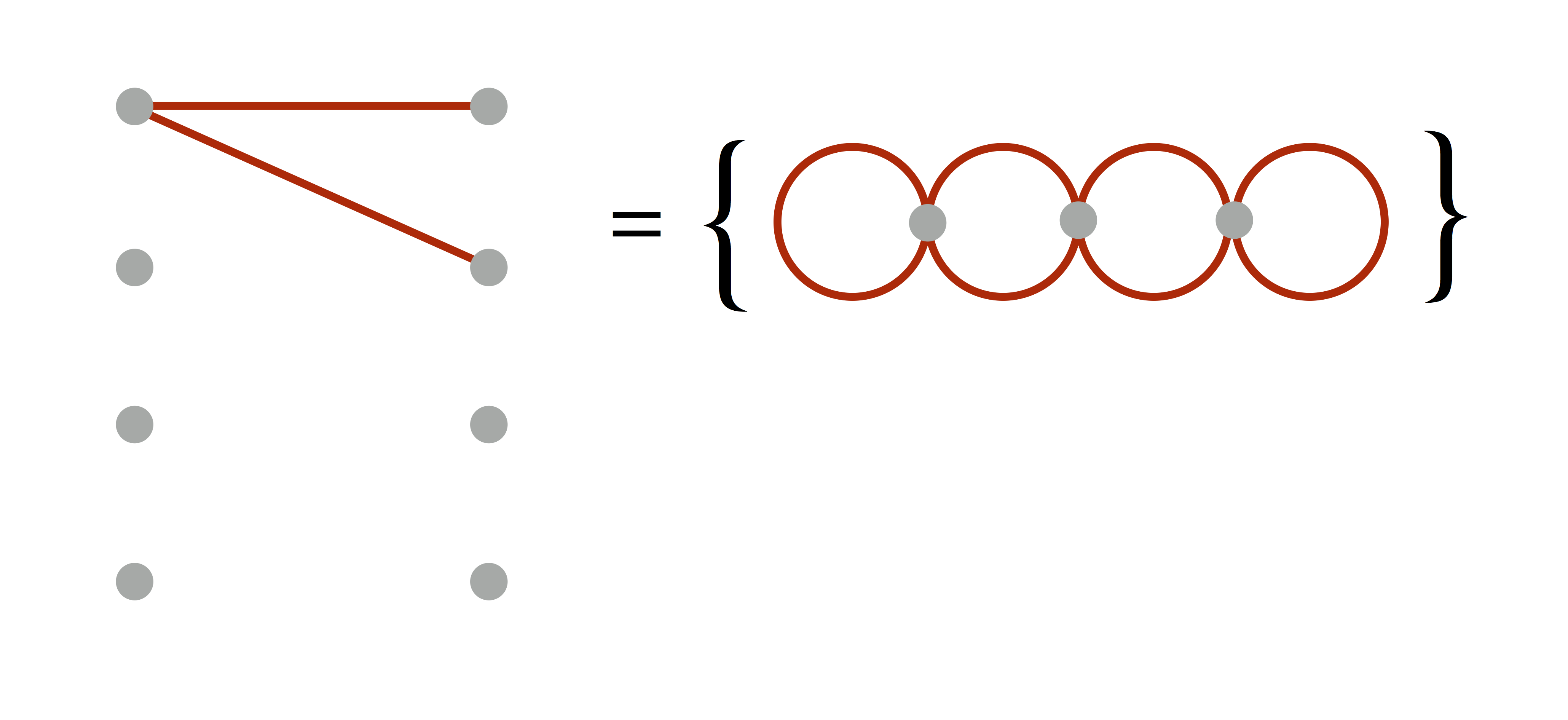} \\
c) \includegraphics[width=7cm,angle=0]{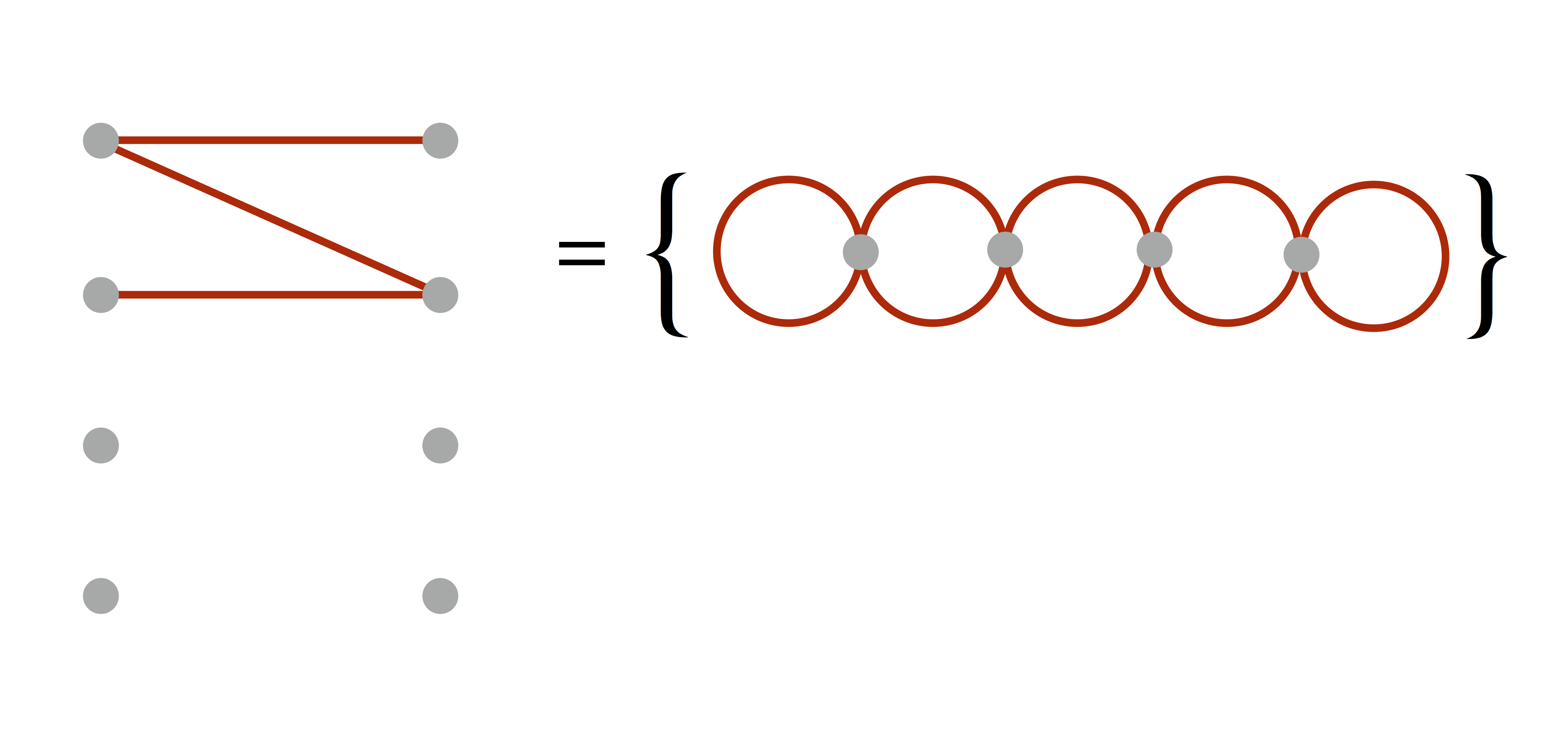} &
d) \includegraphics[width=7cm,angle=0]{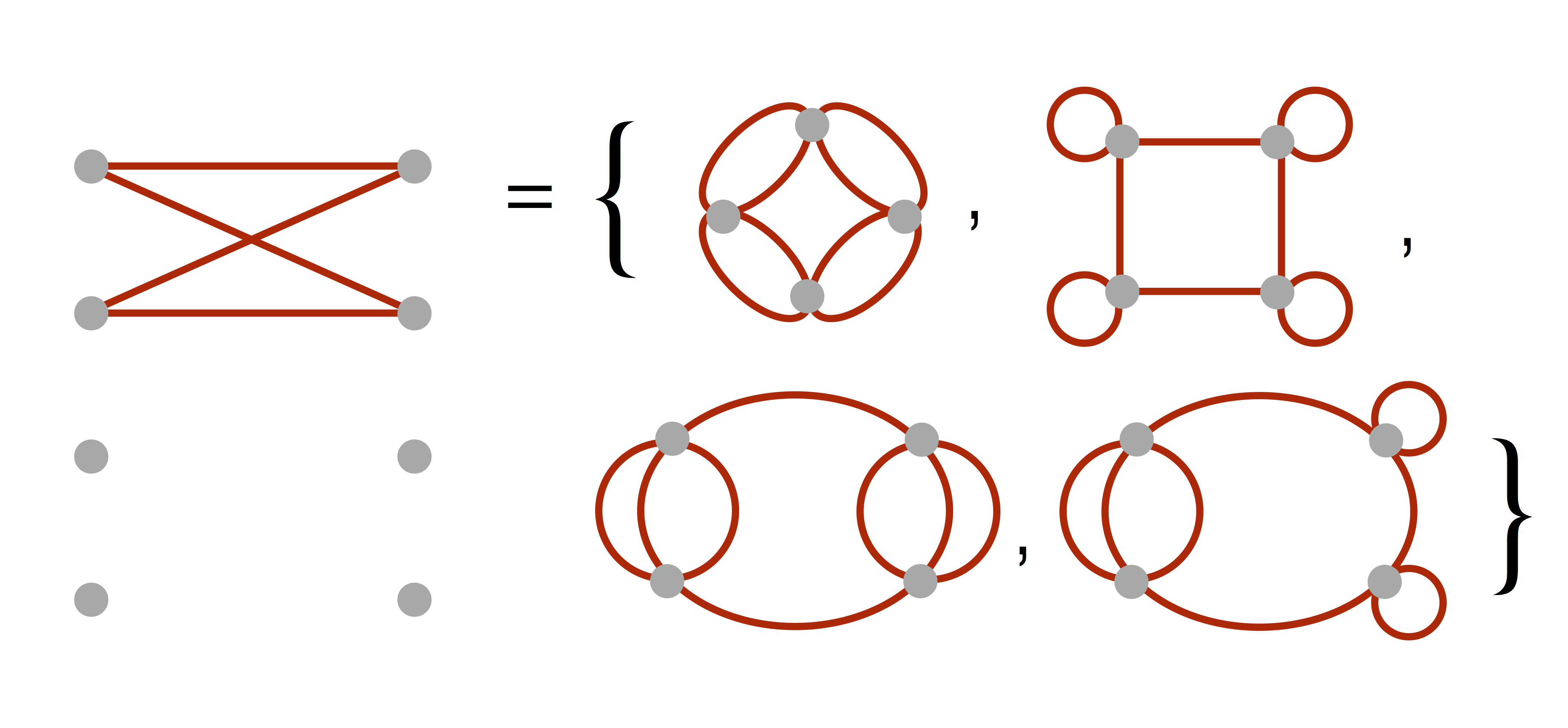}
\end{tabular}
\caption{ Connected spin networks compatible with the Chimera architecture: 
a) $N=2$, b) $N=3$, c) $N=4$, d) $N=4$.} 
\label{V2V3V4}
\end{figure}

Having the definition of a qubit one can try to consider different  spin network topologies which  
are possible to implement directly with the use of Chimera architecture. In Fig. \ref{V2V3V4}
we present connected spin networks with the number of nodes equal to $N=2,3$ and $4$ which can be 
directly embedded into the Chimera graph. 
A single coupling between the qubits in the quantum processor architecture can 
be associated with one or more links in the corresponding spin network. The 
difference between connections can be further encoded in the strength of 
the couplers. In particular in the case a) in Fig. \ref{V2V3V4} there are two possible 
4-valent spin networks which can be associated with two coupled qubits. 
Relating a single coupler with a single link in the spin network is generically
not possible. In the case of 4-valent nodes and a single block of D-Wave 
processor the only possibility is given by the configuration represented in 
blocks b) and c)  in Fig. \ref{Chimera}. The situation corresponds to 
a spin network with $N=8$ qubits and $E=16$ edges forming the Chimera graph. 

As one can notice the structure of Chimera architecture imposes significant 
restrictions on the possible associated spin network topologies. 
In particular, it is not possible to implement a ``triangular'' $N=3$ spin network
directly with the use of elementary qubits. In order to go beyond restrictions of the 
Chimera architecture one can consider effective qubits (chain qubits) composed 
from two or more spins. If the coupling between the qubits is sufficiently negative 
then the qubits will have tendency to align in the same direction, which is 
preferred energetically. In such a case, measurements can be performed on one 
of the elementary qubits contributing to the chain while the remaining qubits can
be considered as ancilla qubits.  
\begin{figure}[ht!]
\centering
\includegraphics[width=4cm,angle=0]{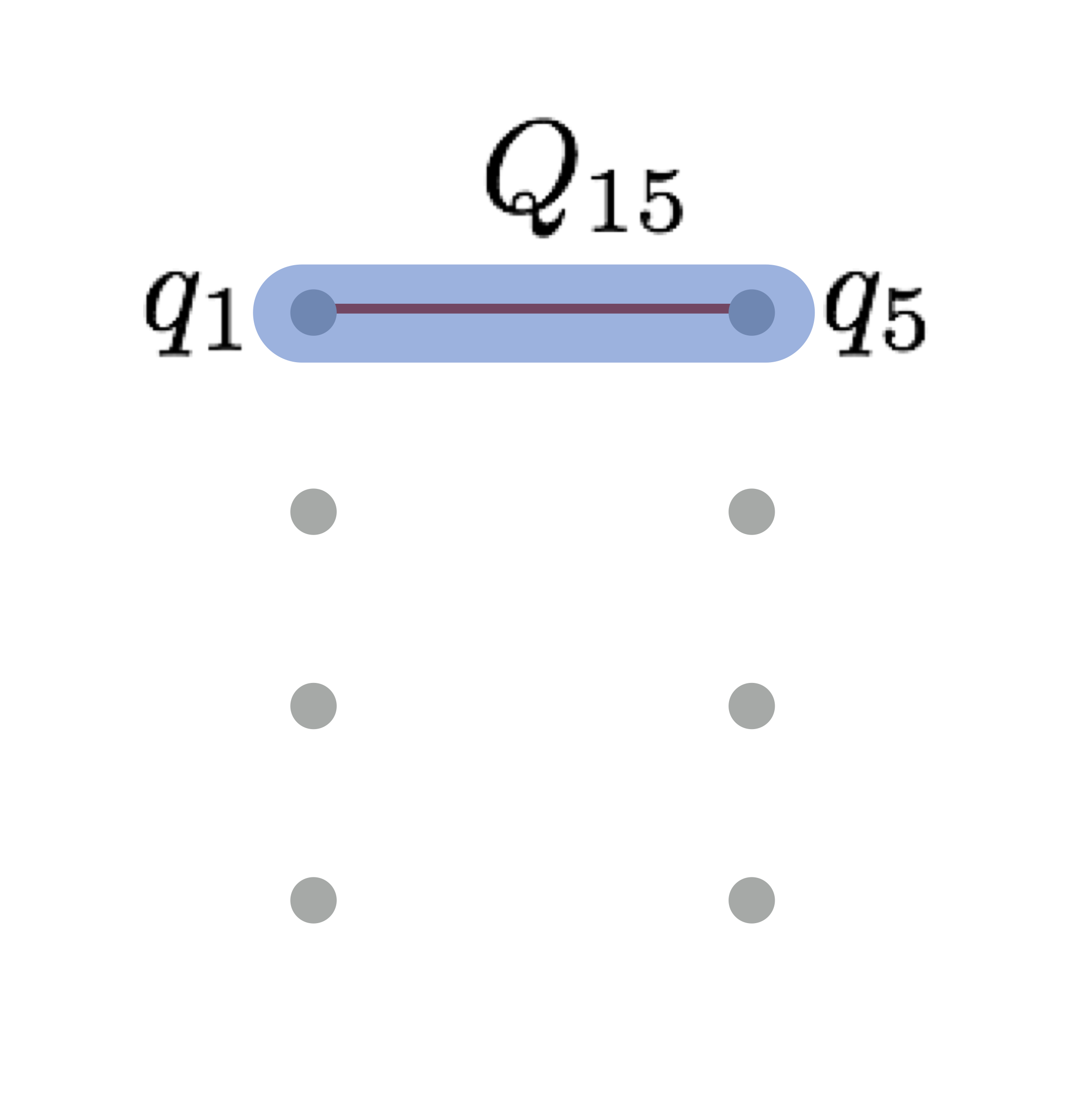}  
\caption{A chain qubit. If the $b_{15}$ coupling is sufficiently negative then the spins 
$q_1$ and $q_5$ will have tendency to align in the same direction.} 
\label{Chain}
\end{figure}

\begin{figure}[ht!]
\centering
\begin{tabular}{cc}
a) \includegraphics[width=8cm,angle=0]{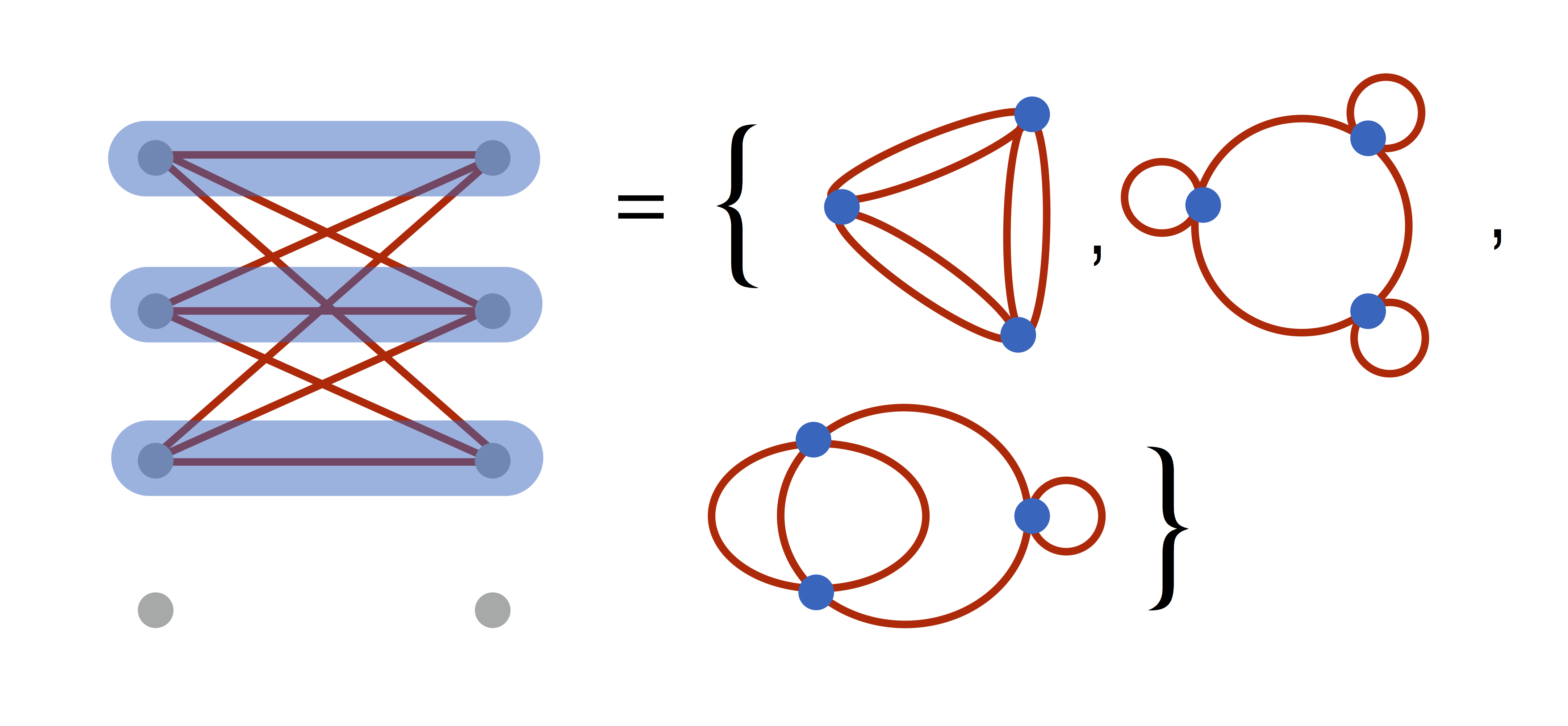}  & 
b) \includegraphics[width=6cm,angle=0]{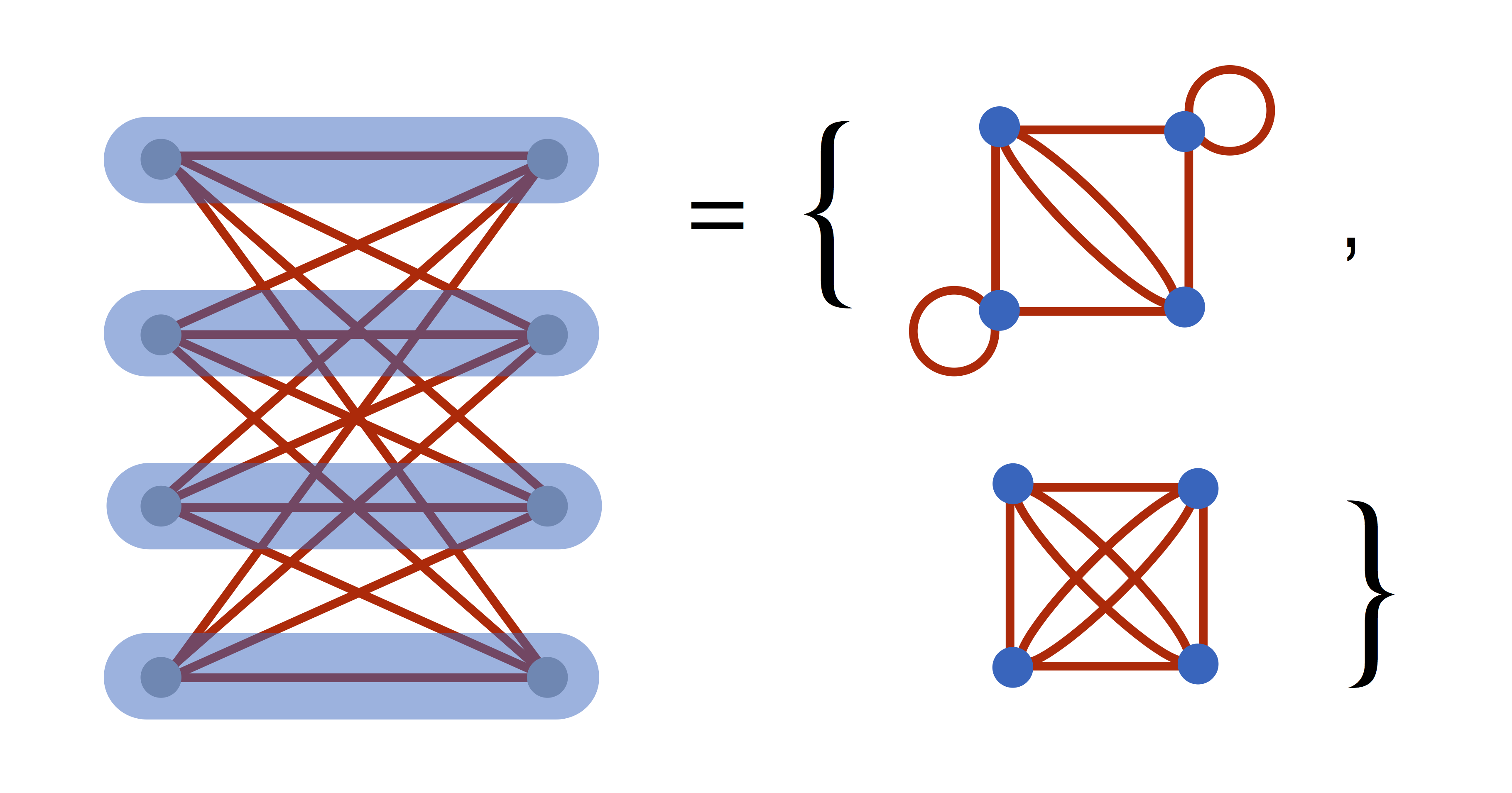}  
\end{tabular}
\caption{Connected spin networks compatible with Chimera architecture: 
a) $N=3$, b) $N=4$.} 
\label{ChainN3N4}
\end{figure}

With the use of chain qubits the dictionary of spin networks can be extended further. 
In particular, previously inaccessible spin networks for $N=3$ and $N=4$ can now be
constructed (see Fig. \ref{ChainN3N4}). 
Worth stressing is that different types of  effective qubits can be considerer 
and Fig. \ref{ChainN3N4} represents only a one of many possible implementations of 
the spin networks under consideration. 

A more extended example is  a regular square lattice with the nearest neighbor connections. 
\begin{figure}[ht!]
\centering
\includegraphics[width=10cm,angle=0]{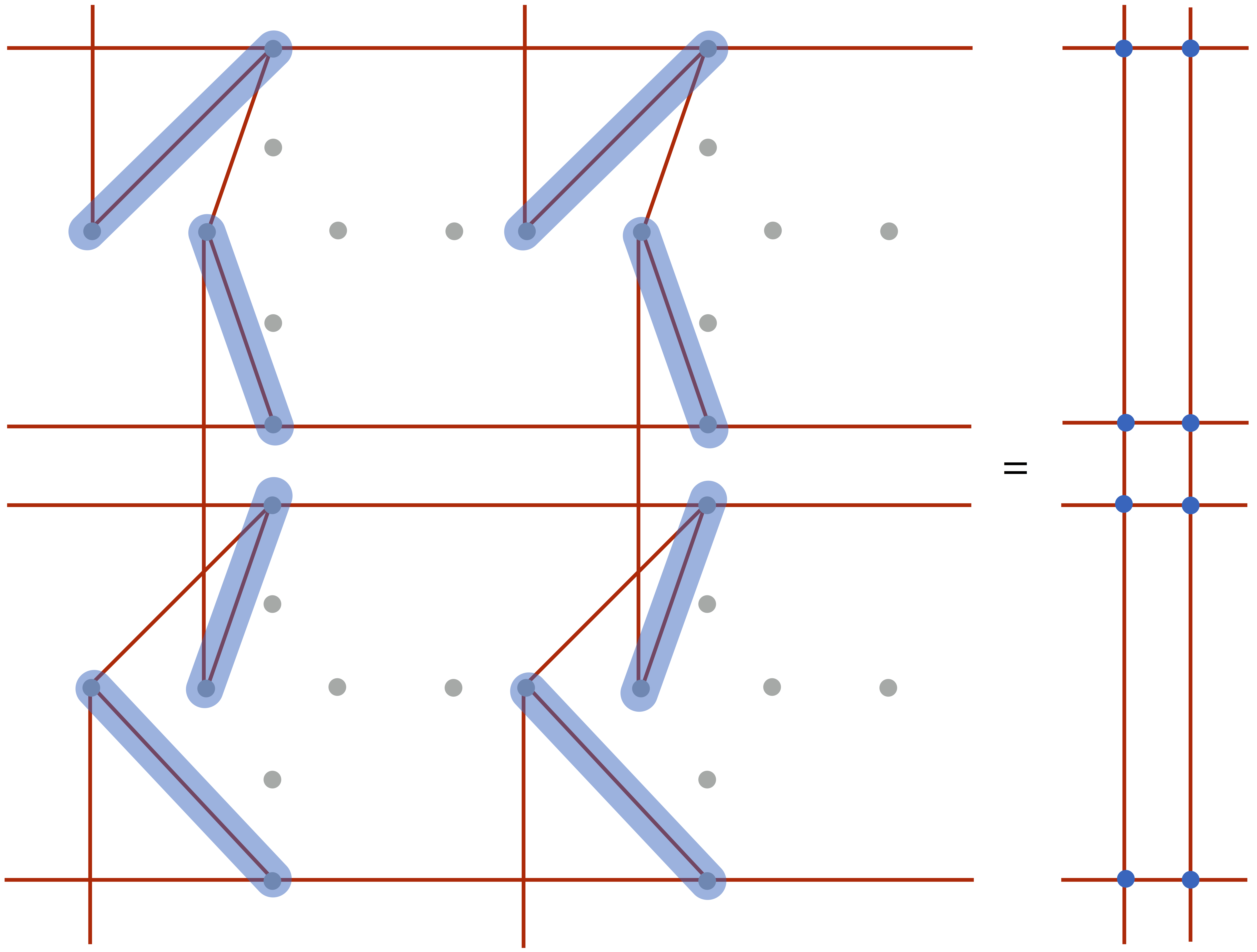}
\caption{Regular lattice spin network embedded with the use of chain qubits. 
Here, four 8 qubit blocks of the D-Wave processor are used.}  
\label{RegVac}
\end{figure}
The regular lattice configuration enables to simulate a 2D \emph{Ising spin 
networks} discussed in Ref. \cite{Feller:2015yta} which provide a toy model of 
extended quantum spacetime. Analysis of such configurations is especially interesting 
in the context of phase transitions and domains formation which may reflect emergence 
of semi-classical spacetime. We will come back to this issue in the next section. 
Furthermore, in the further studies it would be interesting to investigate if the 
3D hexagonal type \emph{Ising spin network} can also be embedded into the architecture 
of the D-Wave processor.  

\section{Simulation of Loop Quantum Gravity} \label{ImpLQG}

The spin network states discussed in the previous sections satisfy the 
Gauss constraint. The diffeomorphism constraint can be imposed 
by considering equivalence classes under the action of diffeomorphism, 
which in practice means that we equate all graphs with the same topology. 

Finally, the scalar (Hamiltonian) constraint remains, which is the most 
difficult one to satisfy. Finding solution to the Hamiltonian constraint in the 
3+1 D can be perceived as the most difficult open problem in LQG 
\cite{Thiemann:2003zv}. 

Basically, the scalar constraints reflects the fact that the total energy of 
gravitational field is equal to zero. The scalar constraint is in general a graph 
changing operator making implementation of such a constraint  a quite 
difficult task. However, the situations in which the graph structure is preserved 
by the action of the constraint may provide an intermediate step towards
the solution of the full problem. Therefore, considering such a prototype 
graph non-changing scalar constraints is worth considering. The question 
is now whether adiabatic quantum computation may be useful here?

In order to answer to this question, let us observe that finding solutions to 
the classical constraint:
\begin{equation}
C \approx 0, \label{constraint1}
\end{equation}
can be mapped into a problem of minimizing some Hamiltonian. 
Because the quantum annealing algorithm is just searching for minimum 
of the spin Hamiltonian (\ref{HI}), making use of adiabatic computations requires 
association of the minimum of the Hamiltonian with solution of the constraint 
(\ref{constraint1}). The simplest way to do it is to consider the Hamiltonian 
$H$ in the following form:
\begin{equation}
H \propto C^2.   \label{Master}
\end{equation} 
In such a case the Hamiltonian is bounded from below and at the ground state 
the constrain (\ref{constraint1}) is naturally satisfied. 

Solution of the constraint (\ref{constraint1}) allows to extract physical 
states and construct a physical phase space $\Gamma_{\rm phys}$ 
(or physical Hilbert space $H_{\rm phys}$) being a subset of 
kinematical phase space $\Gamma_{\rm kin}$\footnote{Here, we define the 
kinematical phase space such that is obtained by solving Gauss and 
diffeomorphism constraints. This corresponds to all possible spin configurations at the 
nodes of a given 4-valent spin network. In the quantum theory, the kinematical 
Hilbert space $H_{\rm kin}$ with respect to the scalar constraint $\hat{C}| \phi \rangle  \approx 0$
is a tensor product of qubit Hilbert spaces defined at $N$ nodes of the 
spin network}. It is important to stress that the minimum energy states 
of the Hamiltonian (\ref{Master}) are just the physical states of the theory and 
they form $\Gamma_{\rm phys}$. If there is no additional matter content, the states
represent also a vacuum gravitational field configuration, described by 
the prototype scalar (Hamiltonian) constraint under considerations. 
Worth mentioning here is that the auxilary Hamiltonian (\ref{Master}) can 
be in some sense considered as a \emph{Master Constraint} introduced 
in LQG, being a square function of constraints (see Ref. \cite{Thiemann:2003zv}). 
    
There are, however, technical limitations in implementation of the procedure
proposed above. This is because, in the D-Wave machine only quadratic 
Hamiltonian functions are allowed. This implies that the scalar constraint 
cannot be of the higher than linear order in the spin variables. On the other hand, 
scalar constraints being of the higher that linear order in the spin variables is 
expected in the full LQG.  

The most general type of the constraint that one consider in the context of 
D-Wave quantum computer is 
\begin{equation}
C = \sum_{i=1}^{N} a_i s_i-c\approx 0,   \label{generalconstraint}
\end{equation} 
with some parameters $a_i,c \in \mathbb{R}$ \footnote{Alternatively, one can 
consider a complex constraint $C = \sum_{i=1}^{N} z_i s_i-c\approx 0$ with 
$z_i,c \in \mathbb{C}$. Then, in order to obtain a real Hamiltonian on has to 
consider $H \propto |C|^2$. This can be extended further to the case of 
multi-constraint model, which we will not discuss here.} and where $s_i\in \{ -1, 1 \} $ 
are classical spin variables. Here, for the sake of simplicity we will consider 
the case with $a_i =1\ \forall i $, such that the prototype scalar constraint (\ref{generalconstraint}) 
takes the following form: 
\begin{equation}
C = \sum_{i=1}^{N}s_i-c\approx 0,  \label{constraint2}
\end{equation}
with some  parameter $c\in \{-N,-N+2, \dots, N-2,N \}$. By squaring 
(\ref{constraint2}) we obtain:
\begin{eqnarray}
C^2=2\left(  \sum_{<i,j>}^N s_{i}s_j+ \sum_{i=1}^{N}(-c)s_i \right)+N+c^2.  
\end{eqnarray} 
Based on this one can propose that the Hamiltonian to be considered is 
\begin{equation}
H = \frac{C^2-N-c^2}{2} =\sum_{<i,j>}^N s_{i}s_j+ h\sum_{i=1}^{N}s_i,  \label{HamiltConst}
\end{equation}
where $h=-c$.  The obtained Hamiltonian corresponds to the QUBO problem with 
a complete graph and equal couplers between the qubits $(b_{ij}=1)$. In this model, 
the ground state (which corresponds to $C=0$) energy is: 
\begin{equation}
H_0 = -\frac{N+c^2}{2}. 
\end{equation}

There is one more important aspect illustrated by the model - a degeneracy of the 
ground state. Namely, there is, in general, no unique spin configurations which is 
minimizing the Hamiltonian (\ref{HamiltConst}). In the model under consideration, 
the vacuum degeneracy depends on both the values of $c$ and $N$. Given the 
$c$ and $N$, determination of the order of degeneracy is a combinatorial problem 
which can be reduced to finding a number of $(N+c)/2$ subset of a set composed of  
$N$ elements, which is given by binomial coefficient: 
\begin{equation}
\left( \begin{array}{c} N \\ \frac{N+c}{2} \end{array} \right). \label{binomial}
\end{equation}
Then, for a fixed N  the maximal degeneracy is obtained for the choices
\begin{equation}
c = 2 \lfloor N/2 \rfloor -N \ \ \vee \ \ c = 2 \lceil N/2 \rceil -N,
\end{equation}
where $ \lfloor x\rfloor$ and $ \lceil x \rceil$ are floor and ceiling functions respectively. 
Based on this, the corresponding maximal degeneracy is equal to 
\begin{equation}
\left(\begin{array}{c} N \\ \lfloor N/2 \rfloor \end{array} \right)
= \left(\begin{array}{c} N \\ \lceil N/2 \rceil \end{array} \right). \label{degeneration}
\end{equation}
The degeneracy is an important quantity because it corresponds to the number of physical 
states satisfying the constraint (\ref{constraint2}).

\begin{figure}[ht!]
\centering
\begin{tabular}{cc}
a) \includegraphics[width=6cm,angle=0]{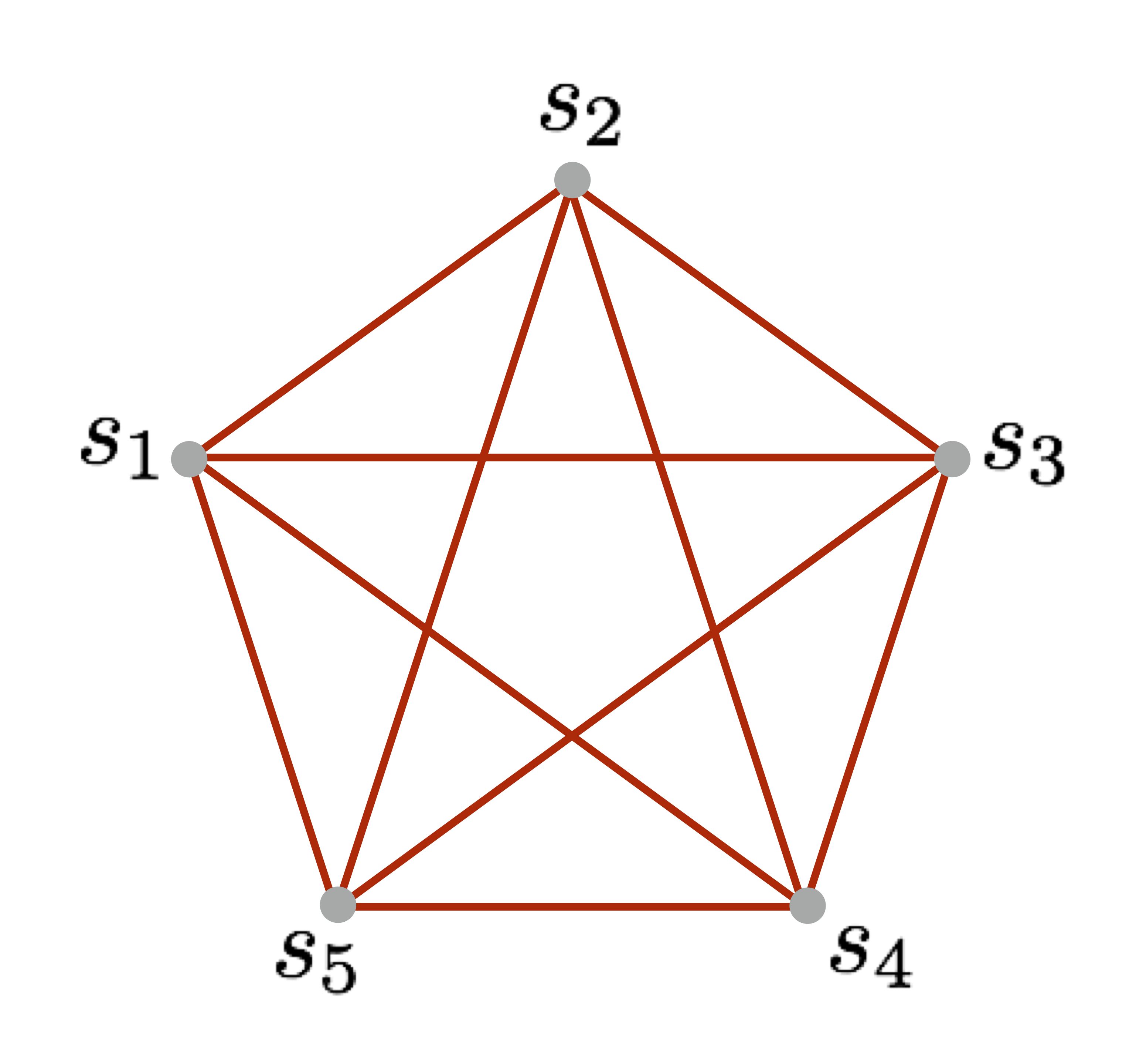} &
b) \includegraphics[width=8cm,angle=0]{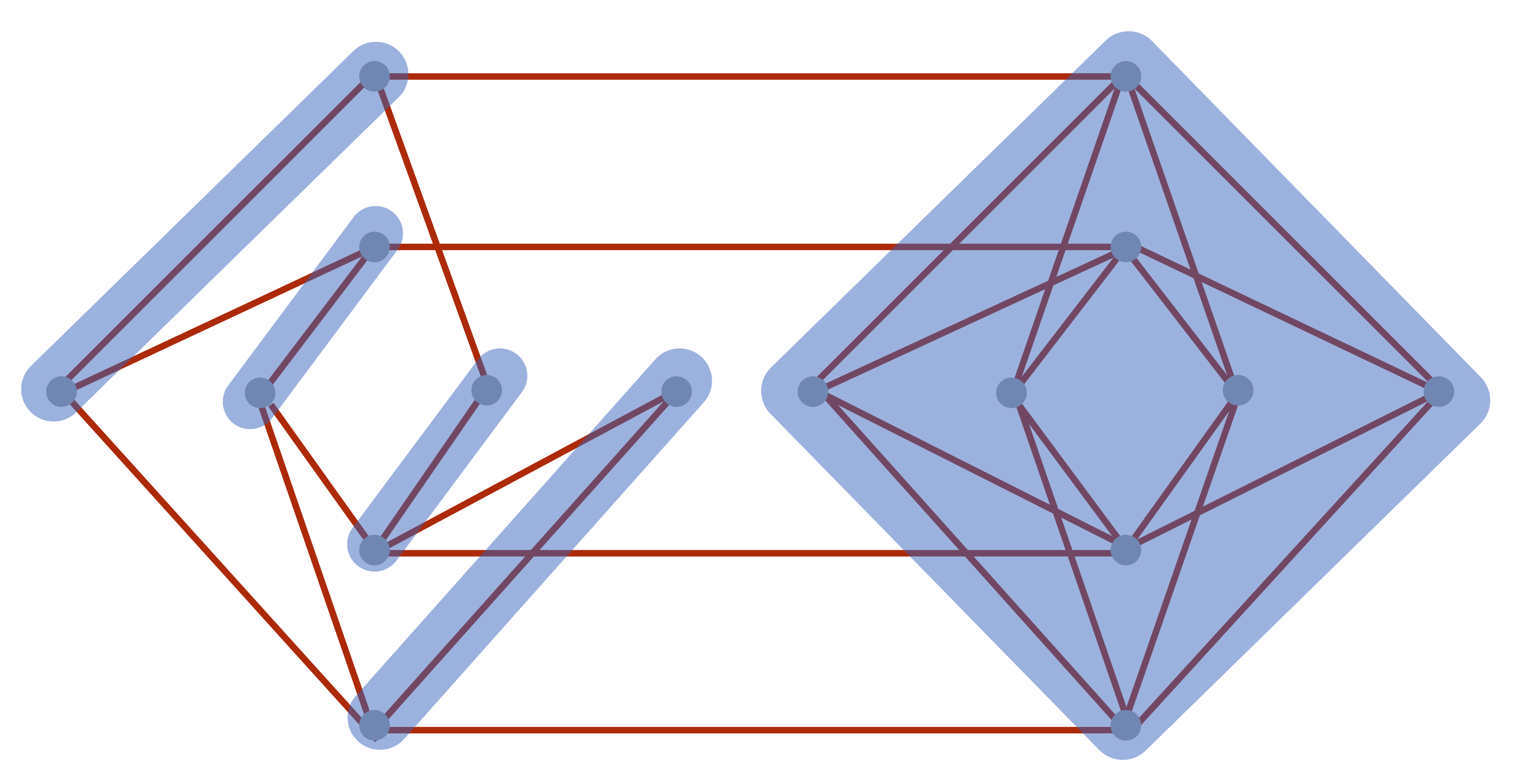} 
\end{tabular}
\caption{a) Pentagram spin network with $N=5$. Each link of the graph is labeled 
with $j=1/2$. The qubits (here modeled by the classical bits $s_i$) are defined at the 
nodes. b) An exemplary embedding of the pentagram spin network on the two neighbor 
blocks of the D-Wave processor. The shadowed regions represent collective (chain) 
qubits which correspond to the nodes of the spin network.} 
\label{Pentagram}
\end{figure}

One can now make relation with the spin networks. For this purpose, let us 
recall that the spin Hamiltonian (\ref{HamiltConst}) corresponding to the 
constraint (\ref{constraint2}) describes a complete graph. Associating 
a spin coupler with a single link of the spin network one can conclude that 
for the 4-valent nodes under considerations the only complete spin network 
must have \emph{pentagram} structure with $N=5$ nodes (see block a) in 
Fig. \ref{Pentagram}) \footnote{The restriction is due to the fact that in the 
considered case all couplers have equal value. Therefore, the couplers 
have to be associated with the same number of links in the spin network. 
The simplest case is when a single coupler is associated with a  single link. 
However, extensions to the other cases are possible if the general form 
of the constraint (\ref{generalconstraint}) is considered.}. 

Such spin network corresponds to geometry of a three-sphere. Furthermore, 
it turns out that introducing composite (chain) spins the pentagram spin network 
can be implemented with the use of two neighbor blocks of the D-Wave processor. 
There are various ways to do so. One of them is presented in block b) in Fig. \ref{Pentagram},
where the shadowed regions correspond to the effective qubits composed our 
of the elementary ones. 

\begin{figure}[ht!]
\centering
\includegraphics[width=11cm,angle=0]{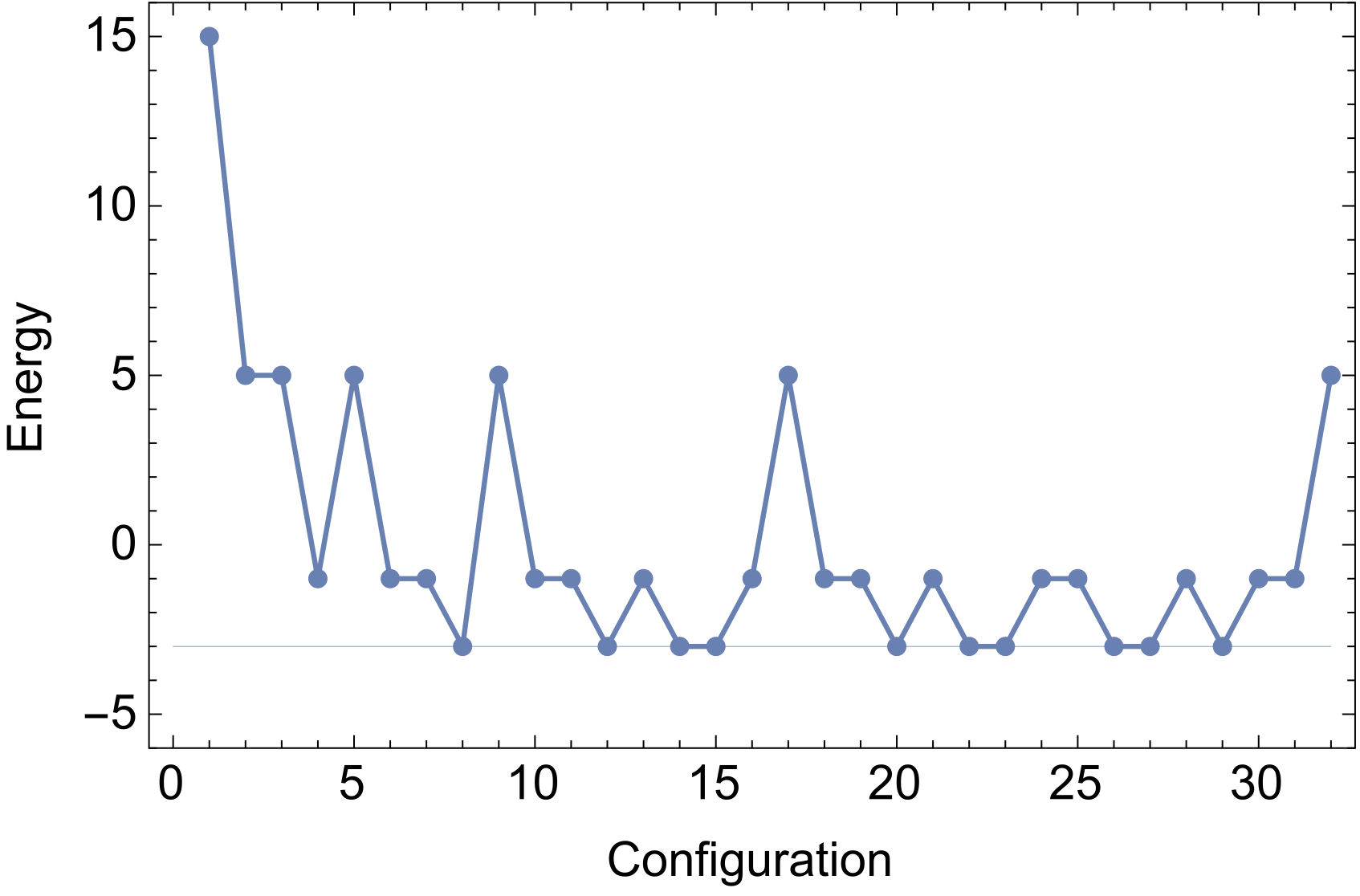}
\caption{Energy landscape of the pentagram spin network with $N=5$ and $c=-1$. 
The ground state configurations satisfy the prototype scalar constraint (\ref{constraint2}) 
and span physical phase space $\Gamma_{\rm phys}$ of the model.} 
\label{Energy}
\end{figure}

Finally, let us take a look at the energy landscape of the model. In Fig. \ref{Energy} 
we plot energies corresponding to all of the spin configurations for $c=-1$. 
The total number of spin configurations corresponds to dimensionality of the 
kinematical space: ${\rm dim}\ \Gamma_{\rm kin} = 2^5=32$. On the other hand, 
the degeneracy of the vacuum of  (\ref{HamiltConst})  gives us dimensionality of 
the physical space ${\rm dim}\ \Gamma_{\rm phys} = \left( \begin{array}{c} 5 \\ \frac{5-1}{2} \end{array} \right) =10$ (here we used Eq. \ref{binomial}). The physical space is a subset of 
kinematical space $\Gamma_{\rm phys}\subset \Gamma_{\rm kin}$ as expected. 

In order to extract all the physical states with the use of adiabatic quantum simulations 
the quantum annealing procedure has to be performed repeatedly. The outcome is a 
superposition of the ground states and the procedure of measurement should select
the particular ground states in the independent runs.  However, as discussed 
in Refs. \cite{Matsuda,Mandra}  the type of quantum annealing procedure used
in the D-Wave quantum computers may not be suited to identify all degenerate ground 
states. The studies suggest that extension beyond the currently employed base 
Hamiltonians is needed to ensure that the ground state manifold is sampled properly. 
Otherwise, the probability of finding some of the possible ground states may be suppressed. 

Assuming that the physical states are selected, analysis of fluctuations of various observables 
is possible to perform. In the case under consideration, one of the interesting possibilities 
would be to investigate volume fluctuations. As we mentioned, the base states corresponding 
to the 4-valent note qubits are eigenstates of the volume operator describing the same 
absolute volume but with the different signs. It is, however, expected that in the classical 
limit only one type of contributions will dominate such that averaged nonvanishing space 
volume will emerge. On the other hand, in a highly quantum state the positive and negative 
contributions can subtract one another leading to the lack of the notion of classical geometry. 
Analysis of correlations of the spins in the physical states could, therefore, say if e.g. 
domains of the same sign of volume form. If yes, that would be a sign of emergence 
of semi-classical spacetime. Furthermore, appearance the long range correlation would 
unavoidably allow to associate a notion of length scale to the spin network configurations. 
Such observation, would be a significant step towards reconstruction of classical spacetime 
directly from the spin network states.  

\section{Summary}

The purpose of this article was to investigate a possibility of implementation of  spin 
networks on the architecture of commercially accessible adiabatic quantum computers. 
In the studies, we focused our attention on spin networks with fixed spin labels ($j=1/2$)
corresponding to fundamental representation of the $SU(2)$ group. In such a case the 
4-valent nodes give rise to two dimensional intertwiner space which was associated with the 
qubit Hilbert space. In the geometric picture, the 4-valent nodes of spin network are dual to 
the 3D simplices and the qubit bases states represent different orientations of a 3-simplex. 

We have shown that in the considered case it is possible to define spin networks 
on architecture of the D-Wave quantum processor. However, due to topological restrictions 
of the Chimera graph not all spin networks are possible to implement with the use of 
elementary qubits. However, some obstacles can be overcame by introducing effective (chain) 
qubits composed our of two or more elementary qubits. Such effective qubits allow to 
implement e.g. regular 2D square lattice topology of the nearest neighbor type of 
interaction Ising model. 

Furthermore, we proposed a method of solving scalar (Hamiltonian) constraints with the 
use of quantum annealing. We have shown that in the case of D-Wave quantum 
annealer, a prototype constraint being a linear function of qubit variables is possible 
to solve. The solutions of the constraint (i.e. physical states) are obtained at the vacuum 
of an appropriate Ising type Hamiltonian. The procedure has been theoretically demonstrated 
for the pentagram spin network, which (as we have shown) can be embedded into 
architecture of the D-Wave processor. Therefore, quantum simulations of the 
proposed procedure can be performed.  

Various generalization of the investigated class  of spin networks are to be considered. 
In particular, the situation which is motivated by the semi-classical limit is when all 
spin labels are equal to some $j \gg1/2$, instead of $j=1/2$. In the case of arbitrary $j$ 
the dimension of the intertwiner space  of a single 4-vertex is 
\begin{equation}
{\rm dim\ Inv} (H_{j}\otimes H_{j}\otimes H_{j}\otimes H_{j})=2j+1.
\end{equation}
Generalization to the case of higher than 4 valence of nodes can be also considered. 
In both cases ancillary qubits have to be introduced in an appropriate manner, which 
is a more difficult task or in some cases perhaps even not possible to do. These and 
other issues related with quantum simulations of spin networks, especially in the context 
of LQG, will be the subject of our further studies. 

\ack

Author is supported by the Grant DEC-2014/13/D/ST2/01895 of the National Centre 
of Science. Author would like to thank to Mario Flory for his careful reading of the 
manuscript and helpful comments.

\end{document}